\documentstyle[12pt,psfig]{l-aa}
\def \src {XB\thinspace1323$-$619}
\def \srcb {1SAX{\thinspace}J1324.4$-$6200}
\def \sax {BeppoSAX}
\def \degmark{^\circ}

\def \ergsec{\hbox{erg s$^{-1}$}}
\def \hcm {\hbox {\ifmmode $ atom cm$^{-2}\else atom cm$^{-2}$\fi}}
\def \arcmin {\hbox{$^\prime$}}
\def \arcsec {\hbox{$^{\prime\prime}$}}

\def\approxgt{\mathrel{\hbox{\rlap{\lower.55ex \hbox {$\sim$}}
        \kern-.3em \raise.4ex \hbox{$>$}}}}
\def\approxlt{\mathrel{\hbox{\rlap{\lower.55ex \hbox {$\sim$}}
        \kern-.3em \raise.4ex \hbox{$<$}}}}
\begin{document}

   \thesaurus{6(13.25.5;  
               08.09.2;  
               08.14.1;  
               08.02.1;  
               02.01.2)}  
\title{Discovery of a new 170~s X-ray pulsar \srcb}

\author{L. Angelini\inst{1}\thanks{Universities Space Research Association}
 \and M.J. Church\inst{2}
 \and A.N. Parmar\inst{3}
 \and M. Ba\l uci\'nska-Church\inst{2}
 \and T.~Mineo\inst{4}
}

   \offprints{L. Angelini}
   \institute{Laboratory for High Energy Astrophysics, Code 660.2, 
	      NASA/Goddard Space Flight Center, MD 20771, USA
   \and
	      School of Physics and Astronomy, University of Birmingham,
              Birmingham, B15 2TT, UK
   \and
              Astrophysics Division, Space Science Department of ESA, ESTEC,
              Postbus 299, 2200 AG Noordwijk, The Netherlands
   \and       
              Instituto IFCAI, via La~Malfa 153, 90146 Palermo, Italy
}

\date{Received 27 August 1998/Accepted 24 September 1998 }

\maketitle

\begin{abstract}

We report the discovery with BeppoSAX of a new $170.84 \pm 0.04$~s 
X-ray pulsar, \srcb, 
found serendipitously in the field of the X-ray binary \src\ in 1997
August. The source and periodicity are also detected in archival ASCA 
data taken in 1994.
The X-ray spectrum is modeled by a power-law with 
a photon index of $1.0 \pm 0.4$ and absorption of
$(7.8 \pm ^{2.7}_{1.1}) \times 10^{22}$~\hcm.
The source is located close to the galactic plane and within 3\arcmin\ 
of the direction of the dark cloud DC 306.8+0.6. The measured 
interstellar absorption 
and cloud size imply a distance $>$3.4~kpc. This implies a 1--10~keV
source luminosity of $>$$1.1 \times 10^{34}$~\ergsec\ during the
BeppoSAX observation.
The source is not detected in earlier {\it Einstein} IPC and EXOSAT CMA
observations, most probably due to reduced detector efficency 
and lower sensitivity to highly absorbed sources.
The X-ray properties suggest that \srcb\ is an accreting neutron star 
with a Be star companion.

\end{abstract}


\section{Observations}

\subsection{BeppoSAX}
\begin{figure}
  \centerline{\psfig{figure=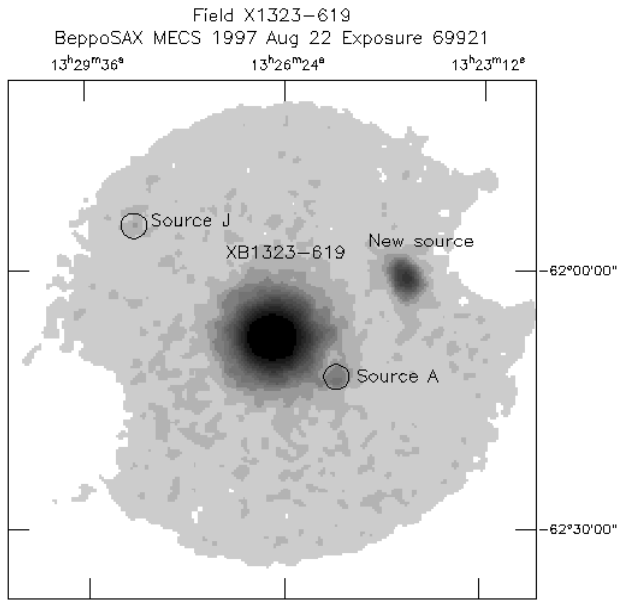,width=8.5cm}}
  \caption[]{The \src\ field MECS image (equinox J2000) smoothed
            with a Gaussian filter with a $\sigma$ of 12\arcsec.
            The two ``cut-outs'' are due to the removal 
            of calibration source events}
  \label{fig:image}
\end{figure}

BeppoSAX (Boella et al. 1997) observed \src, a low-mass X-ray binary 
(Parmar et al. 1998) between 1997 August 22 17:06 and August 24 02:02~UTC.
The Narrow Field Instruments on BeppoSAX 
include the LECS (0.1--10~keV) and three MECS (1.8--10~keV) detectors, 
each at the focus of imaging telescopes.
The fields of view (FOV) of the LECS and MECS are 37\arcmin\ 
and 56\arcmin, respectively.
The exposures in the LECS and MECS are 15 and 70~ks, respectively. 
Data from the MECS 2 and 3 (MECS1 failed in 1997 May) are summed.
The MECS image, shown in Fig.~\ref{fig:image}, reveals in addition to \src, 
the presence of 3 serendipitous sources. Sources
A and J are 2E\thinspace1322.2$-$6157 and 2E\thinspace1325.5$-$6138, 
previously detected with the
{\it Einstein} IPC (Parmar et al. 1989, P89). 
Of interest here is the new source located  $\sim$17\arcmin\ from \src,
with a count rate of $(1.86 \pm 0.07) \times 10^{-2}$~s$^{-1}$. 
The J2000 coordinates, derived from the MECS data, are 
R.A.=$13^{\rm h}\; 24^{\rm m}\; 26\fs3$, 
Dec=$-62\degmark\ 00\arcmin\ 53\arcsec$ 
(galactic ($l,b$)= (306.793, 0.609))
with an uncertainty radius of $1\farcm5$ 
(limited by the current uncertainty in the BeppoSAX 
position reconstruction for 
sources $>$10\arcmin\ off-axis). We designate the source \srcb.
In the LECS the source is 16\farcm7 off-axis. Because of the 
smaller FOV a large fraction of the photons ($\sim$75\%) 
are lost on the detector wall.
This, together with the reduced LECS exposure 
due to observational constraints, prevent the use of these data for
spectral and timing analysis. 

\begin{figure}
    \centerline{\psfig{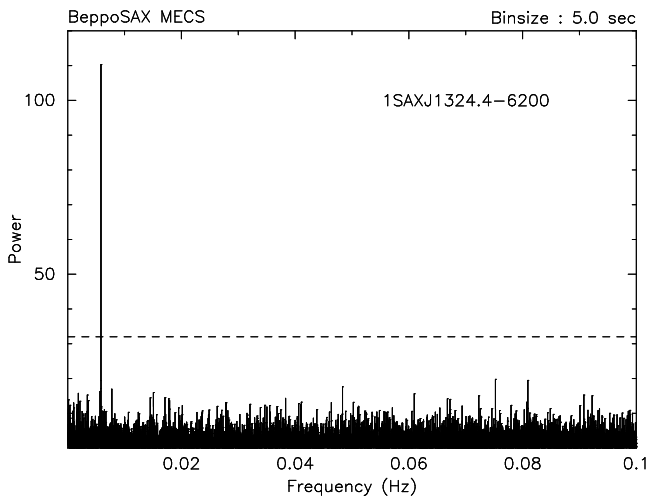}}
    \caption[]{The MECS 1.8--10~keV \srcb\ power spectrum.
    The dashed line indicates the 3$\sigma$ threshold}
    \label{fig:powerspec}
\end{figure}

A total of 1940 MECS events within a 
radius of 4\arcmin\ of \srcb\ were extracted.
The arrival times were corrected to the solar system 
barycenter and binned 
with an integration time of 5~s. A single power spectrum 
(16384 frequencies) was 
calculated for the entire observation and is shown in 
Fig.~\ref{fig:powerspec}. 
A strong peak is detected at $5.853 \times 10^{-3}$~Hz  
(170.85~s) with a significance of $\sim$9$\sigma$. 
No other peaks exceed the $\sim$3$\sigma$ detection threshold. 
The period was refined by cross-correlating pulse profiles each obtained 
by folding data from 12 consecutive intervals.
This yields a pulse period of $170.84 \pm 0.04$~s (at 90\% confidence).
The 1.8--10~keV pulse profile (Fig.~\ref{fig:profile}) 
is approximately sinusoidal with a semi-amplitude (half of the peak to peak modulation 
divided by the mean count rate) of $52 \pm 5$\%. The pulse shape and semi-amplidute
do not show a strong energy dependence. 
The 4.5--10~keV/2.0--4.5~keV hardness ratio is 
constant, except for a slight hardening at $\Phi$ = 0.6--0.7
(where $\Phi$ = 0.0 is the intensity minimum). 
The lightcurve does not show eclipses, dips or strong 
variability with an upper limit of 
$<$14\% {\it rms} at a binning of 400~s.

\begin{figure}[t]
    \centerline{\psfig{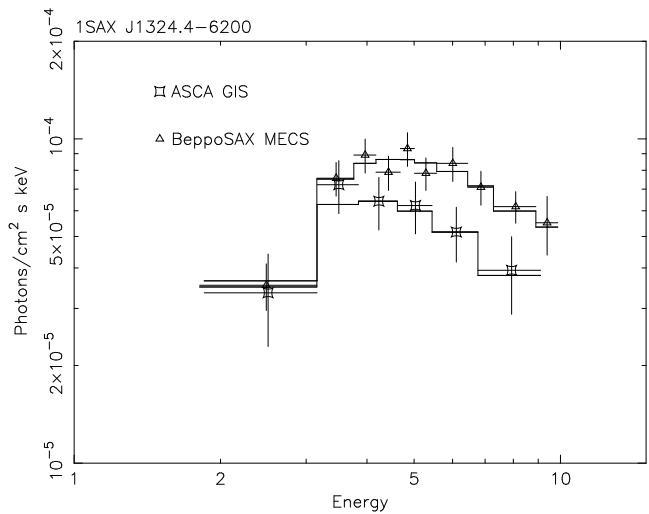}}
    \caption{MECS and GIS unfolded photon spectra. 
             The solid line represents
             the best fit power-law model. Energy is in~keV} 
    \label{fig:spec}
\end{figure}

The MECS spectrum was rebinned to a minimum of 20 counts per bin, 
and was analyzed using an appropriate response matrix for the 
source position in the FOV. 
Due to the close proximity of the galactic ridge emission,
a background spectrum was obtained from the same data using a source
free region and the same extraction radius. 
A power-law model represents 
the data well with a $\chi^2$ of 32 for 44 degrees of freedom (dof), 
with a photon index, $\alpha$, of $1.0 \pm {0.4}$
and absorption, N${\rm _H}$, of $(7.8 \pm ^{2.7}_{1.1}) 
\times 10^{22}$~\hcm\ (Fig.~\ref{fig:spec}).
The 1--10~keV flux is $8.2 \times 10^{-12}$~erg~cm~$^{-2}$~s$^{-1}$.  
No iron K line is detected with a 90\% confidence upper limit of 98~eV to 
the equivalent width of a narrow line at 6.4~keV. 
Blackbody and bremsstrahlung models also fit the data.
For a blackbody model, the temperature, kT, is $2.4 \pm {0.4}$~keV
and N${\rm _{H}} = (4 \pm ^{3}_{2}) \times 10^{22}$~\hcm 
($\chi^2/dof$ = 35/44). For a bremsstrahlung model, the temperature and 
absorption cannot be simultaneously constrained. The 90\% confidence 
limit for kT is $>$10~keV and $( 7.0 <$~N$_{H}$~$< 15.8) \times 10^{22}$~\hcm.
with a $\chi^2/dof \sim 1 $. 
Althought \srcb\ is in the FOV of the non-imaging HPGSPC (5--120~keV) and PDS 
(15--300~keV) detectors, no useful spectral or timing information could be 
extracted from these data. The folded lightcurves are consistent with a 
constant with a semi-amplitude of $<$62 \% (HPGSPC) and $<$10 \% (PDS) respectively. 
The observed counts are dominated by \src\ which is predicted 
to give 10 and 4 times more than \srcb\ in the HPGSPC and PDS respectively. 

\vspace{-0.6cm}

\begin{table}
\caption[]{Observations of \srcb. C is count rate
and $\theta$ off-axis angle.
Upper limits are quoted at 3$\sigma$
confidence. The MECS and GIS count rates are  
uncorrected for factor $\sim$2 vignetting, and $\theta$
is the average 
for each pair of units. The LECS count rate is affected by severe
vignetting (see text)}
\begin{flushleft}
\begin{tabular}{lllll}
\hline\noalign{\smallskip}
Date       & Instrument  & Exp. & C             & $\theta$ \\
(yr mn dy)&             &  (ks)& ($10^{-3}$~s$^{-1}$)& (\arcmin) \\
\noalign{\smallskip\hrule\smallskip}
83 Aug 20 & {\it Einstein} IPC&  4.5 & $<$7.5       & 23.0   \\
84 Feb 11 & EXOSAT CMA        &  6.5 & $<$3.3       &  9.5   \\
85 Feb 11 & EXOSAT CMA        & 93.7 & $<$0.23      &  10.7   \\
94 Aug 04 & ASCA GIS(2,3)     & 19.4 & $22 \pm 3$   & 11.6   \\
97 Aug 22 & SAX LECS     & 14.9 & $5.1 \pm 1.3$& 16.7 \\
97 Aug 22 & SAX MECS(2,3)& 69.9 & $18.7\pm 0.7$& 15.5 \\
\noalign{\smallskip}
\hline
\end{tabular}
\end{flushleft}
\label{tab:sources}
\end{table}

\vspace{-1.3cm}
\subsection{ASCA} 

ASCA (Tanaka et al. 1994) observed \src\ in 1994 August for a total of 20~ks.
\srcb\ is detected in the GIS2 
and GIS3 (0.8--10~keV) instruments (FOV $\sim$45\arcmin) 
at 12\farcm9 and 10\farcm3\ off-axis, respectively.
It is outside the 11\arcmin\ (1 CCD mode) FOV
of the SIS.
The J2000 ASCA position is R.A.=$13 ^{\rm h}\; 24^{\rm m}\; 27\fs6$, 
Dec=$-62 \degmark\ 01\arcmin\ 41\arcsec$ 
(with an uncertainty radius of 1\farcm5), 
consistent with the MECS position. 
Photon event lists and spectra extracted from GIS2 and GIS3 
were combined. The events were binned with an integration time 
of 5~s after barycentric correction. Power spectra of 4096 
frequencies do not reveal significant peaks at the frequency of interest, 
with a limiting semi-amplitude $<$45\%. 
A peak with a significance of $3.3 \sigma$ is found at $\sim$170~s by 
searching with a folding technique in a small range of periods (155--185~s).
\begin{figure}[t]
  \centerline{\psfig{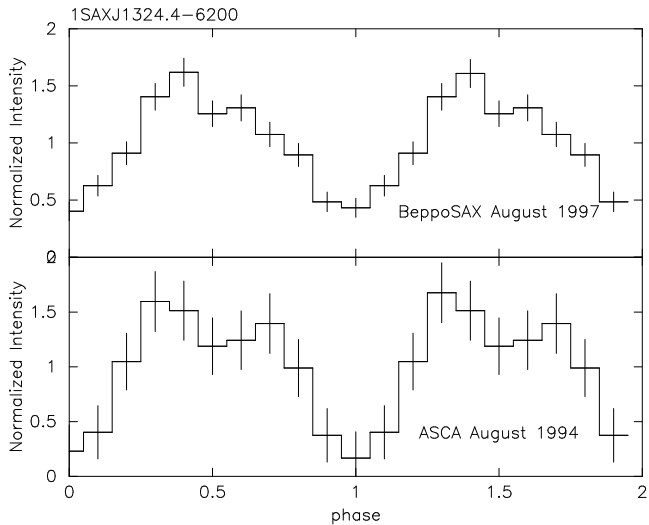}}
  \caption[]{Folded lightcurves at the best period for  
            the BeppoSAX MECS (upper panel, 1.8--10~keV) and 
            ASCA GIS (lower panel, 1.0--10~keV) data. Two cycles 
            are shown}
   \label{fig:profile}
\end{figure}
Cross-correlating the mean pulses obtained from 5 intervals of the ASCA 
data gives a period of $170.35 \pm 0.48$~s.
The 1.0--10~keV GIS pulse profile 
is similar to that obtained with the MECS (Fig.~\ref{fig:profile}). 
A lower limit of the $|\dot P| < 1.0 \times 10^{-8}$~s~s$^{-1}$ 
is obtained combining the ASCA and \sax\ period measurements
(using the lowest value allowed for the 
ASCA measurement).
The lightcurve does not show significant variations with 
an upper limit of 
$<$15\% {\it rms} at a binning of 400~s.

The combined GIS2 and GIS3 source spectrum was rebinned to have
a minimum of 20 counts per bin.
A power-law fit, using the appropriate off-axis response gives
$\alpha = 1.3 \pm _{0.75}^{1.1}$ and 
N$_{H}= (6.6 \pm 2.5) \times 10^{22}$~\hcm\ 
with a $\chi^{2}$ of 27 for 34 dof, 
similar to the values obtained with the MECS. 
The 1--10~keV flux is $5.5 \times 10^{-12}$~erg~cm~$^{-2}$~s$^{-1}$,  
$\sim$30\% less than observed by BeppoSAX.

\vspace{-0.3cm}

\subsection{Earlier observations}  

Table~1 lists the \srcb\ count rates or upper limits
for all available observations.
The source was within the FOV of observations made with the {\it
Einstein} IPC (0.4--4.5~keV) in 1983 and with the EXOSAT CMA (0.04--2~keV) 
in 1984 and 1985 (P89). It is not detected in any of these observations. 
The high absorption measured in the BeppoSAX and ASCA
spectra means that the EXOSAT CMA non-detections are
almost certainly due to the instrument's lower band pass and sensitivity. 
The predicted on-axis count rate of $3\times 10^{-6}$~s$^{-1}$ is 
well below the upper limits given in Table~1.  
In the {\it Einstein} IPC observation,
the pulsar was located at large off-axis angle, very close to a detector rib, 
at reduced detector efficency. Assuming the BeppoSAX
spectral parameters, the expected on-axis count rate is $7.3\times
10^{-3}$~s$^{-1}$, consistent with the upper limit.
The source is not
included in the ROSAT all sky survey catalog (Voges et al. 1996), but the
detection limit of 0.05 count~s$^{-1}$ is above the expected count 
rate of $5 \times 10 ^{-4}$~s$^{-1}$.  
No counterpart was found by searching in the
SIMBAD and HEASARC catalogs. 
\srcb\ lies in a very crowded region of the galactic plane.
The digitized sky survey image contains more than 20 stars in the 
\srcb\ 1\farcm5
radius position uncertainty circle, where the 10 brightest have a Vmag $\ge$13. 
These are unlikely to
be the optical counterpart because of the high N${\rm_{H}}$ 
inferred from the X-ray
spectrum.

\vspace{-0.2cm}

\section{Source location}

\srcb\ lies within 3\arcmin\ of the center of the dark cloud DC
306.8+0.6 (Hartley et al. 1986) which has an angular size of 6\arcmin.  
If \srcb\ is located behind the cloud a limit on the 
N${\rm _{H}}$ and distance can be derived.  
Rowan-Robinson et al. (1991) present a 
simple relation to
equate the IRAS 100 $\mu$m flux to the visual extinction, with an accuracy of 
$\pm$30\% for 90\% of sky.  From the IRAS map at 100~$\mu$m,  
the flux at the source location is 725~MJy~sr$^{-1}$ which translates (using
the above relation) 
to N${\rm _{H} =8.5 \times 10^{22}}$~\hcm.  
This value is
consistent with the N${\rm_{H}}$ measured from the X-ray spectra, 
giving a strong
indication that the absorption measured in \srcb\ is not local to the system,
but due to the dark cloud.  Assuming the lower limit of the 
derived absorption and the average cloud density, as given 
in St\"{u}we (1990) of
3000 cm$^{-3}$ for a single cloud, a cloud size 
of $\sim$6~pc is derived. This
is intermediate between the average cloud size for 
dark cloud complexes
(10~pc) and for simple dark clouds (1pc, St\"{u}we 1990). 
From the angular size of
DC 306.8+0.6 and the above derived size, the 
distance to \srcb\ is estimated to be
$>$3.4~kpc, giving unabsorbed 1--10~keV luminosities of $>$$1.1 \times
10^{34}$ and $>$$7.5 \times 10^{33}$~erg~s$^{-1}$ for the BeppoSAX and ASCA
observations, respectively. 

\vspace{-0.2cm}

\section{Discussion}

The period and spectral properties of \srcb\ are common to both 
accreting magnetized neutron star X-ray pulsars (XRP) and 
accreting magnetized white 
dwarfs in intermediate polar (IP) systems. 
The limit on ${\rm |\dot P| < 1.0 \times 10^{-8}}$~s~s$^{-1}$ from these 
observations is insufficient to distinguish between neutron star 
and white dwarf models.
If \srcb\ is an IP system in, or behind, the dark cloud DC 306.8+0.6, its 
luminosity is above the average of the short 
spin period (33--206~s) systems (Patterson 1994).
This discrepancy could be resolved, if the 
measured absorption is local 
(other IP systems show high N${\rm _{H}}$) and the system closer. 
Patterson (1994) shows that if spin equilibrium is assumed for IP systems,
then the magnetic moment and spin period are correlated. 
Assuming that the measured period is at equilibrium for a magnetic 
moment of $\mu =2 \times 10^{32}$~G~cm$^{-3}$, the luminosity derived  
using the Patterson (1994) relation is $1.3 \times 10^{34}$~erg~s$^{-1}$.
This is above the range
of $10^{31}-4\times 10^{33}$~erg~s$^{-1}$ expected for IP systems.
Some IP such as GK~Per (Watson et al. 1985) show outbursts with luminosities 
similar to 
\srcb. However, it seems unlikely that the source was in outburst for 
the three years covered by the ASCA and BeppoSAX observations.

If \srcb\ is behind the dark cloud the measured luminosity is instead 
consistent with
the XRP luminosity distribution ($10^{32}-10^{38}$ ~erg~s$^{-1}$).
An equilibrium period of 170~s implies a magnetic field of 
$>$$3 \times 10^{12}$~G (Bildsten et al. 1997), which is within the expected 
range for XRP.
The population of accretion-powered XRP has greatly increased,
due to observations with the more sensitive detectors on 
ROSAT, ASCA, RXTE and BeppoSAX. Bildsten et al. (1997) 
list 44 X-ray pulsars and at least 6 others have been 
recently discovered (Israel et al. 1998; Kinugasa et al 
1998; Corbet et al 1998; Marshall et al. 1998; 
Wijnands \& van der Klis 1998;
Hulleman et al. 1998). 
Most XRP are high-mass X-ray binaries (HMXRB) with 
five exceptions (4U\thinspace1626$-$67,
Her X-1, GX1+4, GRO{\thinspace}J1744$-$28, and SAX{\thinspace}J1808.4$-$3658), 
which are low-mass systems.
\srcb\ is unlikely to be one of the exceptions since they are more luminous 
and show different types of variability such as flares, bursting and 
transient behavior.

Roche-lobe filling (Cen X-3) and fed-wind supergiant (Vela X-1) HMXRB are 
the most luminous ($\sim$$10^{35}-10^{38}$~erg~s$^{-1}$) XRPs. 
They shows marked X-ray intensity
variability, due either to eclipses or to 
inhomogeneities in the companion's wind. Unless both X-ray observations 
took place during unusually low states of \srcb, a supergiant companion seems 
unlikely because of the low luminosity and lack of variability.
More than half of the HMXRB are associated with Be star companions and 
typically show transient behavior. This profusion of Be/X-ray pulsar 
systems is unsurprising. Van Paradjis \& McClintock (1995) predict
a ratio $\sim$100 of Be/X-ray binaries to HMXRB with evolved 
companions. Therefore many more Be/X-ray 
binaries are likely to be discovered in galactic plane surveys by 
ASCA and BeppoSAX.
The luminosity of Be-star systems during outbursts can change dramatically 
from $10^{33}$ to $10^{38}$~erg~s$^{-1}$, whereas persistent Be system 
such as X Per, or Be systems in quiescence, have more modest luminosities 
of $\sim$$10^{33}-10^{35}$~erg~s$^{-1}$, similar to \srcb. 
X Per (e.g. Schlegel et al. 1993) and \srcb\ have similar overall spectral
shapes and neither show strong energy-dependent
pulse profiles nor evidence for Fe K line emission.
Be/X-ray systems display a correlation between their spin and orbital periods 
(Corbet 1986; Bildsten et al. 1997) which implies an orbital period of  
$\approxgt$100~days for \srcb.
While \srcb\ is more likely to host a neutron star than a white dwarf,
a change in the spin period of ${\rm |P\dot/P| \sim 10^{-4}}$~yr$^{-1}$ 
(similar to X Per, Haberl 1994), expected because of the lower moment 
of inertia of a neutron star, is needed to confirm this.


\begin{acknowledgements}
The BeppoSAX satellite is a joint Italian-Dutch programme.
We thank the staff of the \sax\ SDC for help with 
these observations. This research has made use of the HEASARC archive, provided 
by NASA's GSFC. 
\end{acknowledgements}


\vspace{-0.4cm}

\begin{thebibliography}{}

\bibitem[1997]{}
Boella G., Butler R.C., Perola G.C., et al., 1997, A\&AS 122, 299

\bibitem[1997]{}
Bildsten L.,  Chakrabarty D., Chiu J., et al., 1997, ApJS 113, 367 

\bibitem[1998]{}
Corbet R.H.D., 1986, MNRAS 220, 1047

\bibitem[1998]{}
Corbet R.H.D., Marshall F.E., Lochner J.C., et al., 1998, IAU Circ. 6803

\bibitem[1986]{}
Haberl F., 1994, A\&A 283, 175

\bibitem[1986]{}
Hartley M., Manchester R.N., Smith R.M., et al., 1986, A\&AS 63, 27

\bibitem[1998]{}
Hulleman F., in 't Zand J.J.M., Heise J., 1998, A\&A 337, L21

\bibitem[1998]{}
Israel G.L., Angelini L., Stella L., et al.,  1998, MNRAS, 298, 502.

\bibitem[1998]{}
Kinugasa K., Torii K., Hashimoto Y., et al., 1998, ApJ 495, 435

\bibitem[1998]{}
Marshall F., Gotthelf E.V., Zhang W., et al., 1998,  ApJ 499, L179

\bibitem[1998]{}
Parmar A.N., Gottwald M., van der Klis M., van Paradijs J., 1989,
ApJ 338, 1024

\bibitem[1986]{}
Parmar A.N., Ba\l uci\'nska-Church M., Church M.J. et al., 1998, in preparation

\bibitem[1994]{}
Patterson J., 1994, PASP 106, 209

\bibitem[1991]{}
Rowan-Robinson M., Hughes J., Jones M., et al., 1991, MNRAS 249, 729

\bibitem[1990]{}
Schlegel E.M., Serlemitsos P.J., Jahoda K., et al., 1993 ApJ 407, 744

\bibitem[1990]{}
St\"{u}we J.A., 1990, A\&A 237, 178

\bibitem[1994]{}
Tanaka Y., Inoue H., Holt S.S., 1994, PASP 46, L37

\bibitem[1995]{}
Van Paradijs J., McClintock J.E., 1995, In: 
Lewin W.H.G., van den Heuvel E.P.J.,
van Paradijs J. (eds.) X-ray Binaries. Cambridge Astrophysics Series 26, p.~1

\bibitem[1994]{}
Voges W., Aschenbach B., Boller Th., et al., 1996, IAU Circ. 6420

\bibitem[1998]{}
Watson M.G., King A.R., Osborne J., 1985, MNRAS 212, 917

\bibitem[1998]{}
Wijnands R., van der Klis M., 1998, Nat 394, 344

\end{thebibliography}
\end{document}